\documentclass[reqno,11pt]{amsart} 
\usepackage[top=1.5in,right=1.25in,left=1.25in,bottom=1.5in]{geometry}
\usepackage{graphicx}
\usepackage{color}
\usepackage{amsmath, amssymb, amsthm}
\usepackage{mathtools}

\definecolor{red}{rgb}{0.7,0,0}
\definecolor{grey}{RGB}{112,112,112}
\definecolor{blue}{RGB}{034,113,179}
\usepackage[colorlinks=true,citecolor=blue,linkcolor=red,urlcolor=grey,backref=page]{hyperref}

\newcommand{\koniec}{\begin{flushright}  $\Box $ \end{flushright}}
\newtheorem{theo}{Theorem}[section] 
\newtheorem{prop}[theo]{Proposition}

\theoremstyle{remark}

\newcounter{mnotecount}[section]

\renewcommand{\themnotecount}{\thesection.\arabic{mnotecount}}

\newcommand{\mnote}[1]
{\protect{\stepcounter{mnotecount}}$^{\mbox{\footnotesize
$
\bullet$\themnotecount}}$ \marginpar{
\raggedright\tiny\em
$\!\!\!\!\!\!\,\bullet$\themnotecount: #1} }

\newcommand{\R}{\mathbb{R}}

\newcommand{\cO}{\mathcal{O}}
\newcommand{\Real}{\mathbb{R}\mathrm{e}}

\def\p{\partial}
\def\be{\begin{equation}}

\def\ee{\end{equation}}

\def\bea{\begin{eqnarray}}
\def\eea{\end{eqnarray}}

\numberwithin{equation}{section}
\begin{document} \date{\today}
\title{The $\phi^4$ kink on a wormhole spacetime}
\author{Alice Waterhouse}

\begin{abstract}
The soliton resolution conjecture states that solutions to solitonic equations with generic initial data should, after some non--linear behaviour, eventually resolve into a finite number of solitons plus a radiative term. This conjecture is intimately tied to soliton stability, which has been investigated for a number of solitonic equations, including that of $\phi^4$ theory on $\mathbb{R}^{1,1}$. We study a modification of this theory on a $3+1$ dimensional wormhole spacetime which has a spherical throat of radius $a$, with a focus on the stability properties of the modified kink. In particular, we prove that the modified kink is linearly stable, and compare its discrete spectrum to that of the $\phi^4$ kink on $\mathbb{R}^{1,1}$. We also study the resonant coupling between the discrete modes and the continuous spectrum for small but non--linear perturbations. Some numerical and analytical evidence for asymptotic stability is presented for the range of $a$ where the kink has exactly one discrete mode.
\end{abstract}

\maketitle

\section{Introduction: the $\phi^4$ kink on $\mathbb{R}^{1,1}$}

One dimensional $\phi^4$ theory is well--documented in the literature (see for example \cite{Manton&Sutcliffe}). The aim of this section is to introduce some notation and some ideas about stability which will be useful when we come to consider the modified theory.

The action takes the form
\be
\nonumber
S=\int_\R\bigg( \frac{1}{2}\eta^{ab}\p_a\phi\p_b\phi + \frac{1}{2}(1-\phi^2)^2\bigg)dx,
\ee
where $x^a=(t,x)$ are coordinates on $\mathbb{R}^{1,1}$ and $\eta^{ab}$ is the Minkowski metric with signature $(-,+)$. Note that the potential has two vacua, given by $\phi=\pm 1$. Finiteness of the associated conserved energy
\[
E = \int_\R \bigg( \frac{1}{2}(\phi_t)^2 + \frac{1}{2}(\phi_x)^2 + \frac{1}{2}(1-\phi^2)^2 \bigg)dx,
\]
requires that the field lies in one of these two vacua in the limits $\phi_\pm=\mathrm{lim}_{x\rightarrow \pm \infty}[\phi(x)]$. We can thus classify finite energy solutions in terms of their topological charge $N=(\phi_+-\phi_-)/2$, which takes values in $\{-1,0,1\}$.


The equations of motion are
\be
\label{eom:R11}
\phi_{tt}=\phi_{xx} + 2\phi(1-\phi^2)
\ee
and we find a static solution $\phi=\mathrm{tanh}(x-c)$ which we call the flat kink. It interpolates between the two vacua and thus has topological charge $N=1$. The constant of integration $c$ can be thought of as the position of the kink. We will henceforth use $\Phi_0$ to denote the static kink at the origin, that is, $\Phi_0(x)=\mathrm{tanh}(x)$. It is evident that no finite energy deformation can affect $N$. For this reason, we say that the kink is \textit{topologically stable}.

\subsection{Linear Stability}

A second notion of stability which will be important to our discussion is linear stability. On discarding non--linear terms, we find that small pertubations $\phi(t,x)=\Phi_0(x)+\mathrm{e}^{i\omega t}v_0(x)$ satisfy the Schr\"odinger equation
\be
\label{eq:zero_order}
L_0v_0:=-v_0^{\prime\prime} - 2(1-3\Phi_0^2)v_0 = \omega^2_0v_0.
\ee
The potential $V_0(x)=-2[1-3\Phi_0(x)^2]$ exhibits a so--called ``mass gap'', meaning that it takes a finite positive value in the limits $x\rightarrow\pm\infty$. In this case, $V_0(\pm\infty)=4$. For $\omega^2>4$, (\ref{eq:zero_order}) admits a continuous spectrum of wave--like solutions.


In addition to its continuum states, the Schr\"odinger operator in (\ref{eq:zero_order}) has two discrete eigenvalues with normalisable solutions given by
\be
\label{eq:flat_vib_modes}
\big(v_0(x),\omega_0\big) = \bigg(\frac{\sqrt{3}}{2}\mathrm{sech}^2(x),0\bigg) \quad \mathrm{and} \quad
\big(v_0(x),\omega_0\big) = \bigg(\frac{\sqrt{3}}{\sqrt{2}}\mathrm{sech}(x)\mathrm{tanh}(x),\sqrt{3}\bigg),
\ee
where we have chosen the normalisation constant such that $\int_{-\infty}^{\infty}v_0^2(x)dx=1$.

The first of these is the zero mode of the kink. Its existence is guaranteed by the translation invariance of (\ref{eom:R11}), and up to a multiplicative constant it is equal to $\Phi_0^\prime(x)$. Excitation of this state corresponds to performing a Lorentz boost. In the non--relativistic limit, this amounts to replacing $\Phi_0(x)$ with $\Phi_0(x-vt)$ for some $v\ll 1$ \cite{Manton&Sutcliffe}.

The second normalisable solution, called an \textit{internal mode} has non--zero frequency $\omega$, and is thus time periodic. In the full non--linear theory, it decays through resonant coupling to the continuous spectrum \cite{Manton&Merabet}. This phenomenon is of considerable interest in non--linear PDEs, and was studied in a more general setting in \cite{SW98}. The corresponding process in the modified theory will be discussed in section \ref{sec:dynamics}.

Linear stability of the kink is equivalent to the Schr\"odinger operator $L_0$ in (\ref{eq:zero_order}) having no negative eigenvalues, so that linearised perturbations cannot grow exponentially with time. One way to see that the kink is linearly stable is via the Sturm oscillation theorem:

\begin{theo}[Sturm]\label{th:sturm} Let $L$ be a differential operator of the form
\[
L=-\frac{d^2}{dx^2}+V(x)
\]
on the smooth, square integrable functions $u$ on the interval $[0,\infty)$, with the boundary condition $u(0)=0$ (corresponding to even parity) or $u^\prime(0)=0$ (corresponding to odd parity). Let $\omega^2$ be an eigenvalue of $L$ with associated eigenfunction $u(x;\omega)$. Then the number of eigenvalues of $L$ (subject to the appropriate boundary conditions) which are strictly below $\omega^2$ is the number of zeros of $u(x;\omega)$ in $(0,\infty)$.
\end{theo}

Note that the symmetry of (\ref{eq:zero_order}) under $x\mapsto -x$ means that any solution on the interval $[0,\infty)$ has a corresponding solution on the interval $(-\infty,0]$, and these solutions can be pieced together to make a smooth solution on $(-\infty,\infty)$ as long as the boundary conditions at $x=0$ are chosen to ensure parity $\pm1$. Thus there is a one--to--one correspondence between solutions on $[0,\infty)$ and solutions on $(-\infty,\infty)$ which are smooth at $x=0$. Since the eigenfunctions (\ref{eq:flat_vib_modes}) have no zeros on the interval $[0,\infty)$, it follows that there can be no eigenfunctions with $\omega^2<\omega_0^2=0$, and thus the kink is linearly stable.

\subsection{Asymptotic stability}
The final notion of stability that we will consider is that of asymptotic stability. Stated simply, asymptotic stability of the kink means that for sufficiently small initial pertubations, solutions of (\ref{eom:R11}) will converge locally to $\Phi_0(r)$ or its Lorentz boosted counterpart. This was proved in \cite{KowalczykMM} for odd perturbations, but has not been proved in the general case.

Generalisation of the finite energy $\phi^4$ kink to higher dimensional Minkowski spacetimes is prohibited by a scaling argument due to Derrick \cite{Derrick}. In order to construct a higher dimensional $\phi^4$ kink, we must add curvature. In the next section we introduce a curved background, and show that a modified $\phi^4$ kink exists on this background. We will also examine a limit in which the modified kink reduces to the flat kink. In section \ref{sec:linearised_pert} we consider linearised perturbations around the modified kink, proving that it is linearly stable and comparing its discrete spectrum to that of the flat kink. In section \ref{sec:dynamics} we examine the mode of decay to the modified kink in the full non--linear theory, in particular the resonant coupling of its internal modes to the continuous spectrum.

\section{The static kink on a wormhole}

We now replace the flat $\mathbb{R}^{1,1}$ background with a wormhole spacetime $(M,g)$, where
\be
\nonumber
g=-dt^2+dr^2+(r^2+a^2)(d\vartheta^2 + \sin^2\vartheta d\varphi^2)
\ee
for some constant $a>0$, and $-\infty<r<\infty$. This spacetime was first studied by Ellis \cite{Ellis} and Bronnikov \cite{Bronnikov}, and has featured in a number of recent studies about kinks and their stability \cite{wavemaps,SG}. Note the presence of asymptotically flat ends as $r\rightarrow\pm\infty$, connected by a spherical throat of radius $a$ at $r=0$.

Our action is the modified by the presence of a non-flat metric:
\be
\nonumber
S=\int\bigg( \frac{1}{2}g^{ab}\p_a\phi\p_b\phi + \frac{1}{2}(1-\phi^2)^2\bigg)\sqrt{-g}dx,
\ee
where $x^a$ are now local coordinates on $M$. Variation with respect to $\phi$ gives
\be
\label{eom.coordfree}
\square_g\phi+2\phi(1-\phi^2)=0
\ee
where $\square_g\phi=\frac{1}{\sqrt{-g}}\p_a(g^{ab}\sqrt{-g}\p_b\phi)$. We assume $\phi$ is independent of the angular coordinates $(\vartheta,\varphi)$, so (\ref{eom.coordfree}) can be written explicitly as
\be
\label{eom.coords}
\phi_{tt}=\phi_{rr}+\frac{2r}{r^2+a^2}\phi_r+2\phi(1-\phi^2).
\ee

The conserved energy in the theory is given by
\be
\nonumber
E=\int_{-\infty}^{+\infty} \bigg( \frac{1}{2} (\phi_t)^2 + \frac{1}{2} (\phi_r)^2 + \frac{1}{2}(1-\phi^2)^2\bigg)(r^2+a^2)dr,
\ee
which we require to be finite. This imposes the condition $\phi^2\rightarrow 1$ as $r\rightarrow\pm\infty$, so that the field lies at one of the two vacua at both asymptotically flat ends.

Static solutions $\phi(r)$ satisfy
\be
\label{eom.static}
\phi^{\prime\prime}+\frac{2r}{r^2+a^2}\phi^{\prime}=-\frac{d}{d\phi}\bigg(-\frac{1}{2}(1-\phi^2)^2\bigg),
\ee
which, can be thought of as a Newtonian equation of motion for a particle at position $\phi$ moving in a potential $\mathcal{U}(\phi)=-(1-\phi^2)^2/2$, with a time dependent friction term.

In addition to the two vacuum solutions, we have a single soliton solution which interpolates between the saddle points at $(-1,0)$ and $(1,0)$ in the in the $(\phi,\phi^\prime)$ plane. Its existence and uniqueness among odd parity solutions follow from a shooting argument: suppose the particle lies at $\phi=0$ when $r=0$. If its velocity $\phi^\prime(0)$ is too small, it will never reach the local maximum of the potential at $\phi=1$, but if $\phi^\prime(0)$ is too large it will overshoot the maximum so that $\mathcal{U}(\phi)\rightarrow -\infty$ as $r\rightarrow\infty$, thus having infinite energy. Continuity ensures that there is some critical velocity $\phi^\prime(0)$ such that the particle reaches $\phi=1$ in infinite time and has zero velocity upon arrival. This corresponds to the non--trivial kink solution, which we call $\Phi(r)$. Time reversal implies that $\phi\rightarrow -1$ as $r\rightarrow -\infty$, and that the anti--kink $\phi(r)=-\Phi(r)$ is also a solution.

\begin{figure}
\includegraphics[width=14cm]{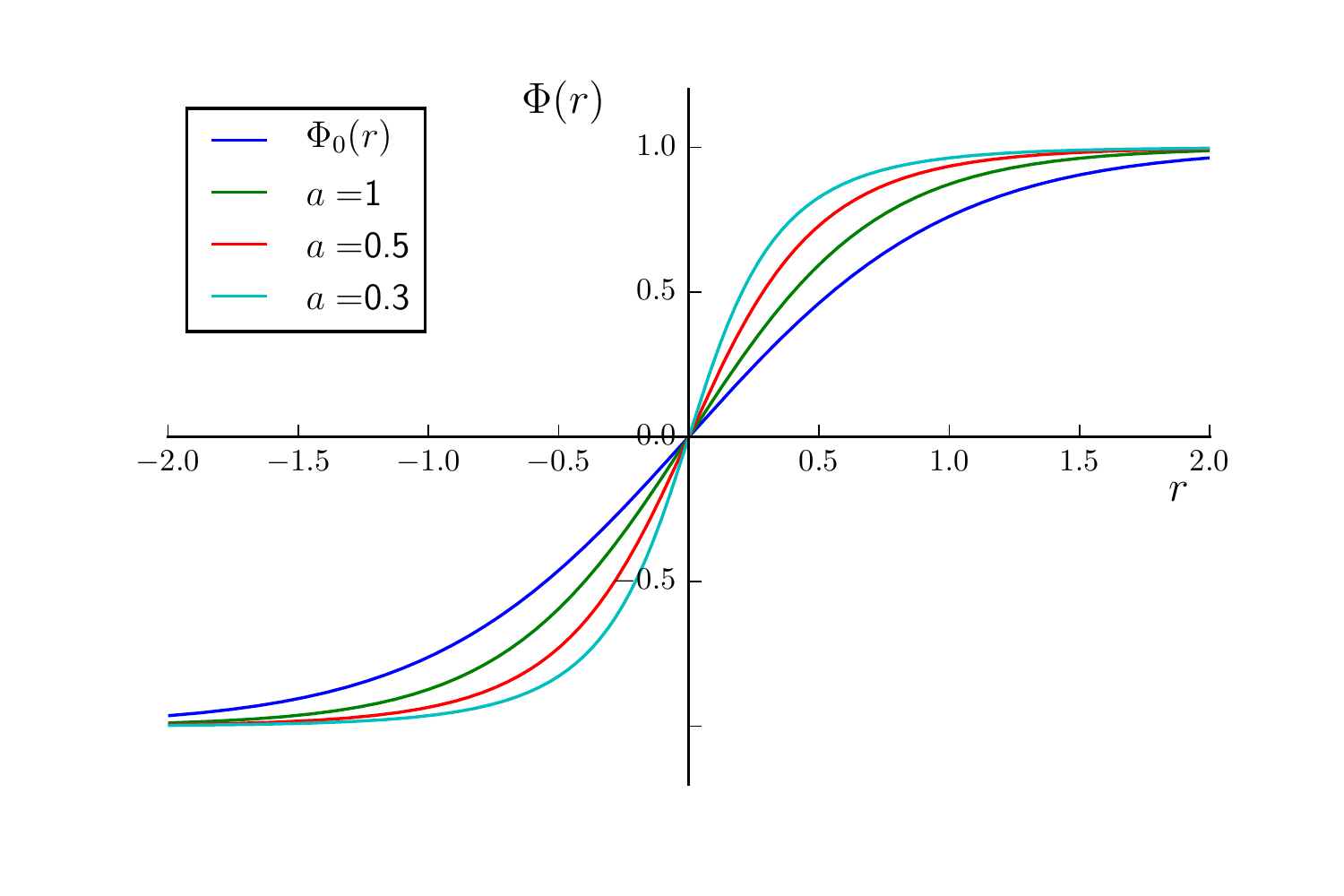}
\caption{\label{fig:several_kinks}The kink solution for several values of $a$, along with the flat kink $\Phi_0(r)$.}
\end{figure}

We can find $\Phi(r)$ numerically using a shooting method for the gradient at $r=0$. Figure \ref{fig:several_kinks} shows such numerically generated kinks for several values of $a$. Note that the absolute value of $\Phi(r)$ is always greater than or equal to that of the flat kink $\Phi_0(r)$, and that at fixed non--zero $r$, the absolute value of $\Phi$ decreases as $a$ increases. The reason for this will become clear in section \ref{sec:linearised_pert}. In section \ref{sec:large_a} we examine $\Phi(r)$ in the limit where $a$ is large, finding that it reduces to the flat kink $\Phi_0(r)$, and examining its departure from the flat kink at first order in $1/a^2$.

We again label the values at the boundary as
\[
\phi_\pm := \lim_{r\rightarrow\pm\infty} \Phi(r)\in \{\pm 1\}.
\]
Since no finite energy deformation can change the value of the topological charge $N=(\phi_+-\phi_-)/2 \in \{-1,0,1\}$, we again conclude that $\Phi(r)$ is topologically stable.

\subsection{Large $a$ limit} \label{sec:large_a}
As $a\rightarrow\infty$, equation (\ref{eom.coords}) becomes the standard equation (\ref{eom:R11}) for the flat kink. It is thus helpful to expand the modified kink in $\epsilon^2:=1/a^2$ for small $\epsilon^2$, since we can then solve both (\ref{eom.coords}) and (\ref{schrod}) analytically up to $\mathcal{O}(\epsilon^4)$. We shall denote the static kink by $\Phi_\epsilon(r)$ in this limit. It satisfies
\be
\label{eq.large_a}
\Phi_\epsilon^{\prime\prime}+\frac{2r\epsilon^2}{\epsilon^2 r^2+1}\Phi_\epsilon^{\prime}=-2\Phi_\epsilon(1-\Phi_\epsilon^2).
\ee

Setting $\Phi_\epsilon(r)=\Phi_0(r)+\epsilon^2\Phi_1(r)+\mathcal{O}(\epsilon^4)$ we obtain at order zero the equation (\ref{eom:R11}) of a static kink on $\mathbb{R}^{1,1}$. This has solution $\Phi_0(r)$, where we have chosen the kink at the origin to restrict to solutions with odd parity.

At order $\epsilon^2$ we find that $\Phi_1(r)$ must satisfy
\[
\Phi_1^{\prime\prime} + 2r\mathrm{sech}^2r=2\Phi_1(2-3\mathrm{sech}^2r).
\]
The unique solution which is odd and decays as $r\rightarrow\pm\infty$ is given by
\[
\Phi_1(r)=\frac{1}{24}\mathrm{sech}^2r(f_1(r)+f_2(r)+f_3(r)),
\]
where
\be
\nonumber
\begin{split}
f_1(r) &=r\big[3-8\mathrm{cosh}(2r)-\mathrm{cosh}(4r)\big], \\
f_2(r) &=\mathrm{sinh}(2r)\big[8\mathrm{log}(2\mathrm{cosh}(r))-1\big] + \mathrm{sinh}(4r)\mathrm{log}(2\mathrm{cosh}(r)), \\
f_3(r) &= \frac{\pi^2}{2} + 6r^2 +6\mathrm{Li}_2(-\mathrm{e}^{-2r}),
\end{split}
\ee
and $\mathrm{Li}_2(z)$ is the dilogarithm function, i.e.  the special case $s=2$ of the polylogarithm
\[
\mathrm{Li}_s(z)=\sum_{k=1}^\infty \frac{z^k}{k^s}.
\]

To show that $\Phi_1(r)$ is odd, note that $\mathrm{sech}^2r$ is an even function, and that $f_1$ and $f_2$ are constructed from products of even and odd functions, and hence are odd. To see that $f_3$ is also odd, we use Landen identity for the dilogarithm:
\be
\nonumber
\begin{split}
\mathrm{Li}_2(-\mathrm{e}^{-2r})+\mathrm{Li}_2(-\mathrm{e}^{2r}) &= -\frac{\pi^2}{6} - \frac{1}{2}\big[\mathrm{log}(\mathrm{e}^{-2r})\big]^2 \\
&= -\frac{\pi^2}{6} - 2r^2,
\end{split}
\ee
thus verifying $f_3(r)+f_3(-r)=0$. 

\begin{figure}
\includegraphics[width=14cm]{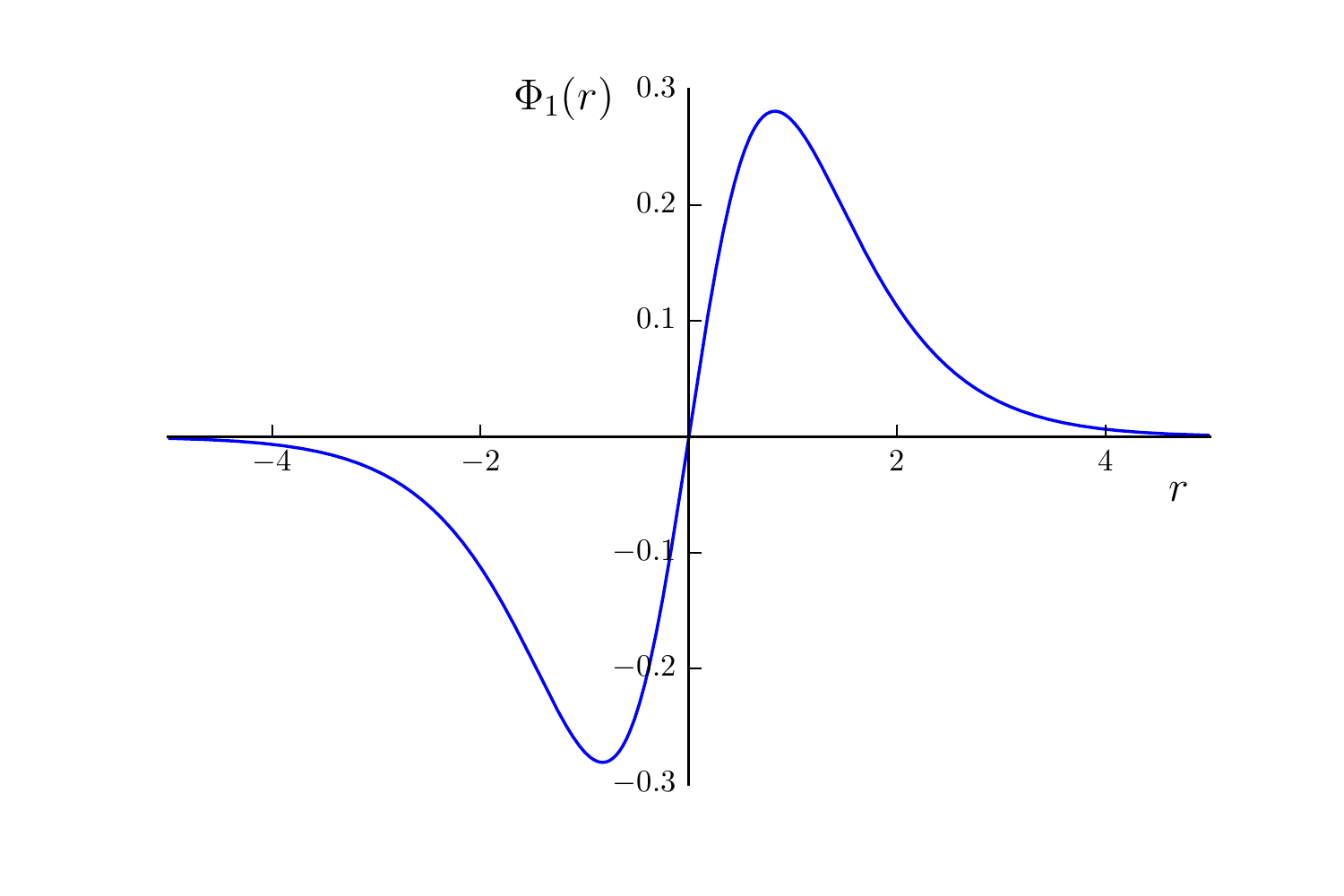}
\caption{\label{fig:Phi_1}The order $\epsilon^2$ perturbation to the static kink on $\mathbb{R}^{1,1}$.}
\end{figure}

We now turn to the behaviour of $\Phi_1(r)$ as $r\rightarrow\infty$. Since $\mathrm{sech}^2r\sim 4\mathrm{e}^{-2r}$ for large $r$, we need only consider terms in the $\{f_i\}$ of order $\mathrm{e}^{2r}$ or higher. We first note that 
\be
\nonumber
\begin{split}
\mathrm{log}(2\mathrm{cosh}r)&=\mathrm{log}(\mathrm{e}^r(1+\mathrm{e}^{-2r}))=r+\mathrm{log}(1+\mathrm{e}^{-2r}) \\
&=r+\mathrm{e}^{-2r} + \mathcal{O}(\mathrm{e}^{-4r}).
\end{split}
\ee
Then
\be
\nonumber
\begin{split}
f_1(r)&=-4r\mathrm{e}^{2r} -\frac{r}{2}\mathrm{e}^{4r}+\mathcal{O}(\mathrm{e}^r) \\
f_2(r)&=\frac{1}{2}\mathrm{e}^{2r}(8r+8\mathrm{e}^{-2r}-1)
+\frac{1}{2}\mathrm{e}^{4r}(r+\mathrm{e}^{-2r})+\mathcal{O}(\mathrm{e}^r) \\
&=4r\mathrm{e}^{2r} +\frac{r}{2}\mathrm{e}^{4r}+\mathcal{O}(\mathrm{e}^r),
\end{split}
\ee 
so $f_1(r)+f_2(r)=\mathcal{O}(\mathrm{e}^r)$. Since $f_3(r)=\mathcal{O}(r^2)$ for large $r$, we see that $\Phi_1(r)$ vanishes as $r\rightarrow\infty$, as we expect. Note that its vanishing as $r\rightarrow-\infty$ then follows using parity. A plot of $\Phi_1(r)$ is shown in figure \ref{fig:Phi_1}.

\section{Linearised perturbations around the kink}
\label{sec:linearised_pert}
To study the linear stability of the kink, we first plug
\be
\label{eq:pert}
\phi(t,r)=\Phi(r)+w(t,r)
\ee
into equation (\ref{eom.coords}), discarding terms non--linear in $w$. Imposing the fact that $\Phi(r)$ satisfies (\ref{eom.static}), we find
\be
\nonumber
w_{tt}=w_{rr}+\frac{2r}{r^2+a^2}w_r+2w(1-3\Phi^2).
\ee
For $w(t,r)=\mathrm{e}^{i\omega t}(r^2+a^2)^{-1/2}v(r)$, this becomes a one--dimensional Schr\"odinger equation
\be
\label{schrod}
Lv:=(-\p_r\p_r+V(r))v=\omega^2 v,
\ee
where the potential is given by
\be
\label{QMpotential}
V(r)=\frac{a^2}{(r^2+a^2)^2}-2(1-3\Phi^2).
\ee

\begin{figure}
\includegraphics[width=14cm]{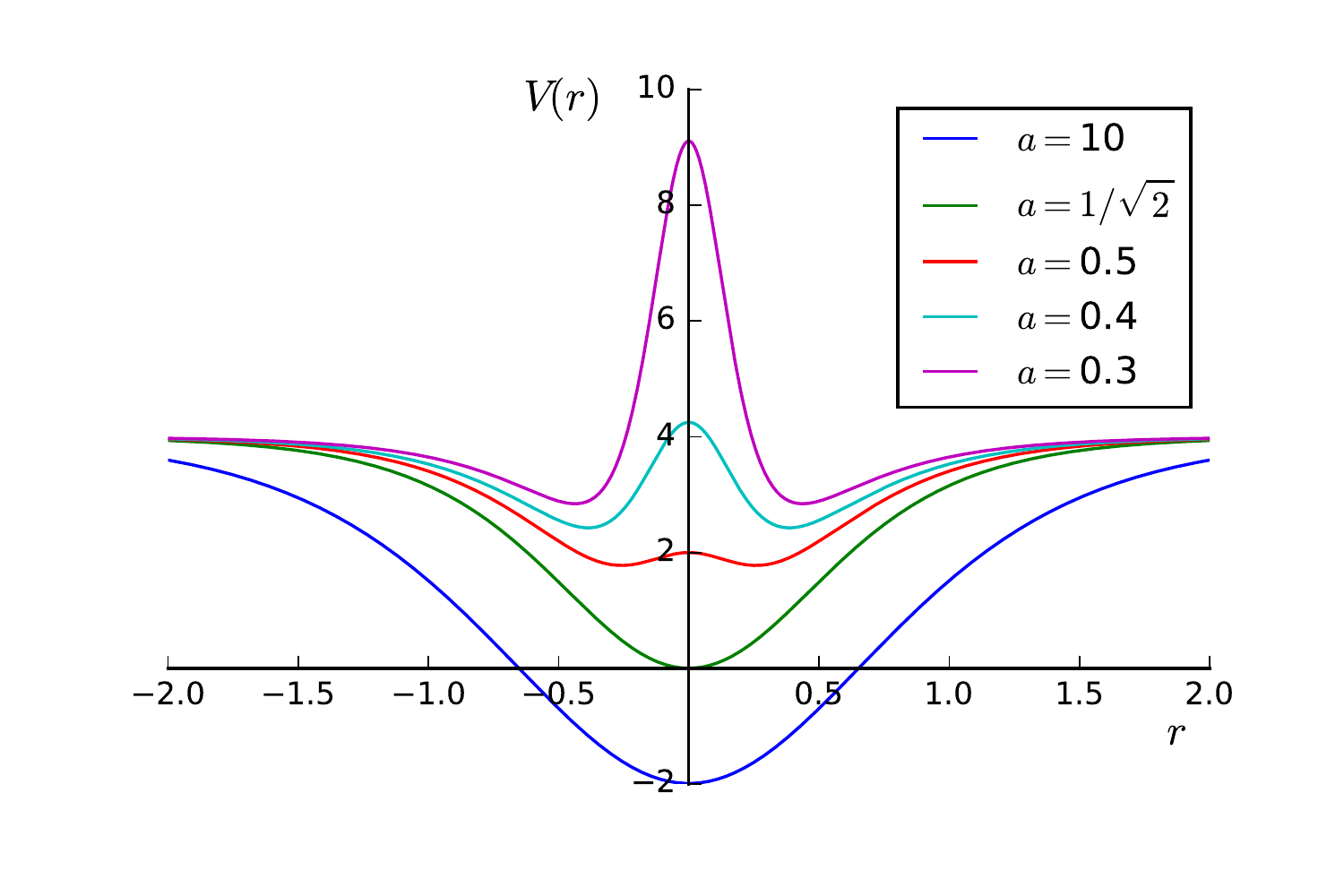}
\caption{\label{fig:several_potentials}The potential of the 1--dimensional quantum mechanics problem arising from the study of stability of the soliton for values of $a$ between $a=10$ and $a=0.3$. In particular, note that those with $a<1/\sqrt{2}$ are everywhere positive.}
\end{figure}

Figure \ref{fig:several_potentials} shows the potential $V(r)$ for several values of $a$. Note that for large $a$ it has a single well with a minimum at $r=0$, close to the potential $V_0$ corresponding to the flat kink. As $a$ decreases, the critical point at $r=0$ becomes a maximum with minima on either side, creating a double well. We find numerically that this happens at about $a=0.55$.

\begin{prop}
The kink solution $\Phi(r)$ is linearly stable.
\end{prop} 

{\bf Proof.} We first decompose the potential $V(r)$ in (\ref{schrod}) as $V=V_0+V_1+V_a$, where
\[
V_0 = -2[1-3\Phi_0(r)^2],\qquad 
V_1 = 6[\Phi(r)^2-\Phi_0(r)^2],\qquad V_a = \frac{a^2}{(r^2+a^2)^2}.
\]
As discussed above, we know that the operator $L_0 = -\p_r\p_r + V_0$ has no negative eigenvalues. It then follows that $L$ itself has no negative eigenvalues as long as the functions $V_1(r)$ and $V_a(r)$ are everywhere non--negative.

The latter is obvious; to prove the former we recall that we can think of $\Phi(r)$ and $\Phi_0(r)$ as the trajectories of particles moving in a potential $\mathcal{U}(\phi)$, where $r$ is imagined as the time coordinate. The particle corresponding to $\Phi(r)$ suffers an increased frictional force compared to $\Phi_0(r)$, i.e.
\be
\label{eq:particles}
\Phi_0^{\prime\prime}=-\frac{\p\mathcal{U}}{\p\phi}\bigg|_{\phi=\Phi_0},\qquad
\Phi^{\prime\prime}+\frac{2r}{r^2+a^2}\Phi^\prime=-\frac{\p\mathcal{U}}{\p\phi}\bigg|_{\phi=\Phi}.
\ee
Both $\Phi$ and $\Phi_0$ interpolate between the maxima of $\mathcal{U}$ at $\phi=\pm 1$; reaching the minimum ($\phi=0$) when $r=0$.

Multiplying the equations (\ref{eq:particles}) by $\Phi_0^\prime$ and $\Phi^\prime$ respectively, then integrating from $r$ to $\infty$, we have that at every instant of time
\be
\label{eq:energy}
\frac{1}{2}(\Phi_0^\prime)^2 + \mathcal{U}(\Phi_0) = 0,\qquad
\frac{1}{2}(\Phi^\prime)^2 + \mathcal{U}(\Phi) = \int_r^\infty\frac{2r}{r^2+a^2}(\Phi^\prime)^2dr.
\ee
These equations are equivalent to conservation of energy for each of the particles. Note that the integral on the RHS is non--negative for $r\geq 0$, and vanishes only at $r=\infty$. In particular, when $r=0$ we have $\mathcal{U}(\Phi)=\mathcal{U}(\Phi_0)=-1/2$, so $\Phi^\prime(0)>\Phi_0^\prime(0)$. This means $V_1(r)$ is initially increasing from zero.

For $V_1(r)$ to return to zero at some finite $r=r_0$, we would need that $\Phi(r_0)=\Phi_0(r_0)$ at a point where $\Phi^\prime(r_0)\leq\Phi_0^\prime(r_0)$. However, this is made impossible by equations (\ref{eq:energy}), since at such a point $\mathcal{U}(\Phi)=\mathcal{U}(\Phi_0)$ and the integral on the RHS is positive. Hence $V_1(r)$ remains non--negative for all $r>0$, and thus for all $r$ since it is even in $r$.
\koniec

\subsection{Finding internal modes numerically}
\label{sec:finding_bound_states_numerically}
Bound states of the potential (\ref{QMpotential}) correspond to internal modes of the kink like the odd solution of (\ref{eq:zero_order}) in (\ref{eq:flat_vib_modes}). In contrast, for frequencies greater than $\omega = 2$, solutions to (\ref{schrod}) are interpreted as radiation. It is possible to search for bound states of (\ref{QMpotential}) numerically by setting $v(0)=1,v^\prime (0)=0$ or $v(0)=0,v^\prime(0)=1$ depending on the required parity and then using a bisection method to search for the value of $\omega^2$ for which $v(r)$ goes to zero at the boundaries. This procedure will only be effective within the range of $r$ for which $\Phi(r)$ is calculated.

\begin{figure}
\includegraphics[width=14cm]{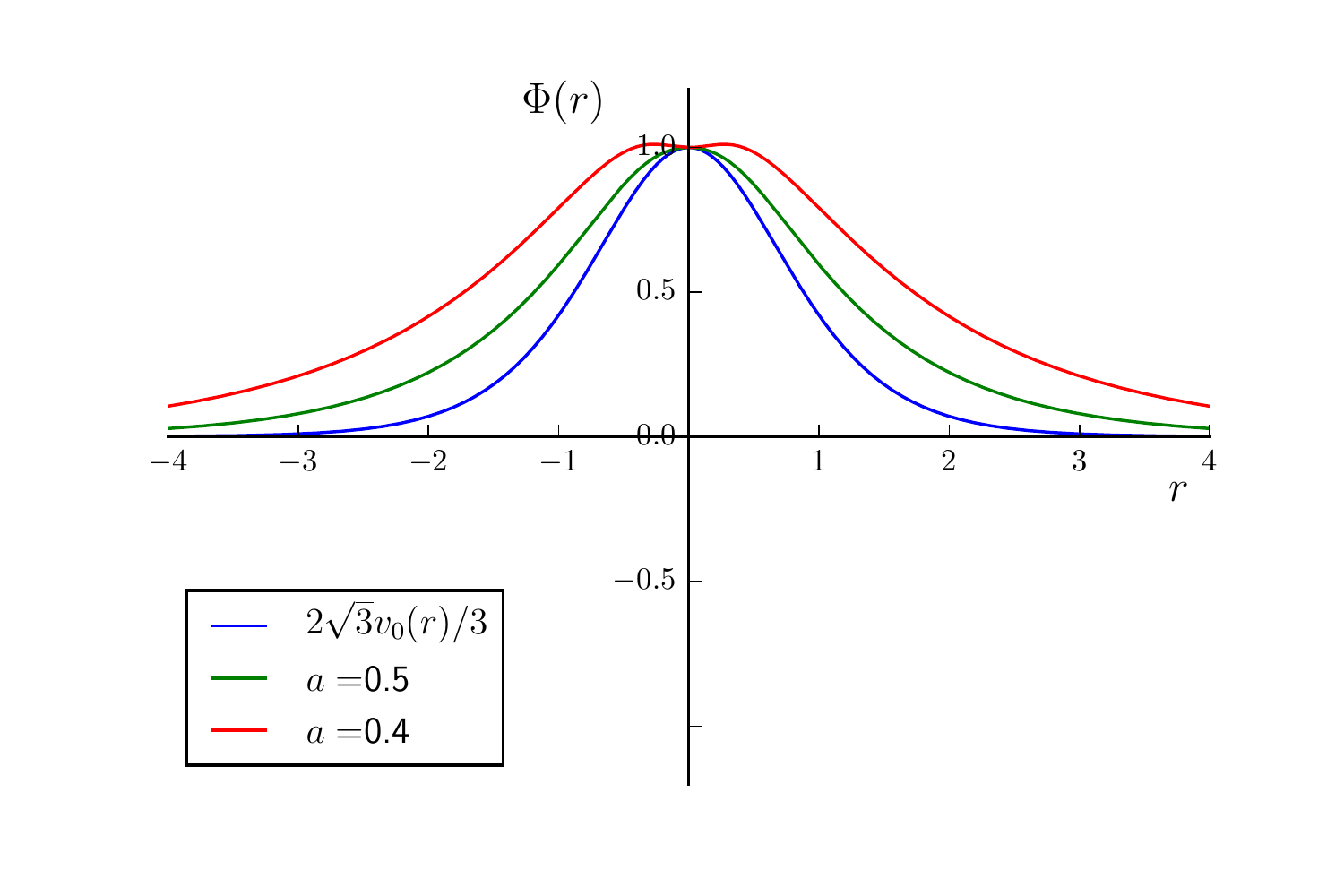}
\caption{\label{fig:even_states} Even bound states of the potential $V_0(r)$ and of the potential $V(r)$ for two different values of $a$.}
\end{figure}

\begin{figure}
\includegraphics[width=14cm]{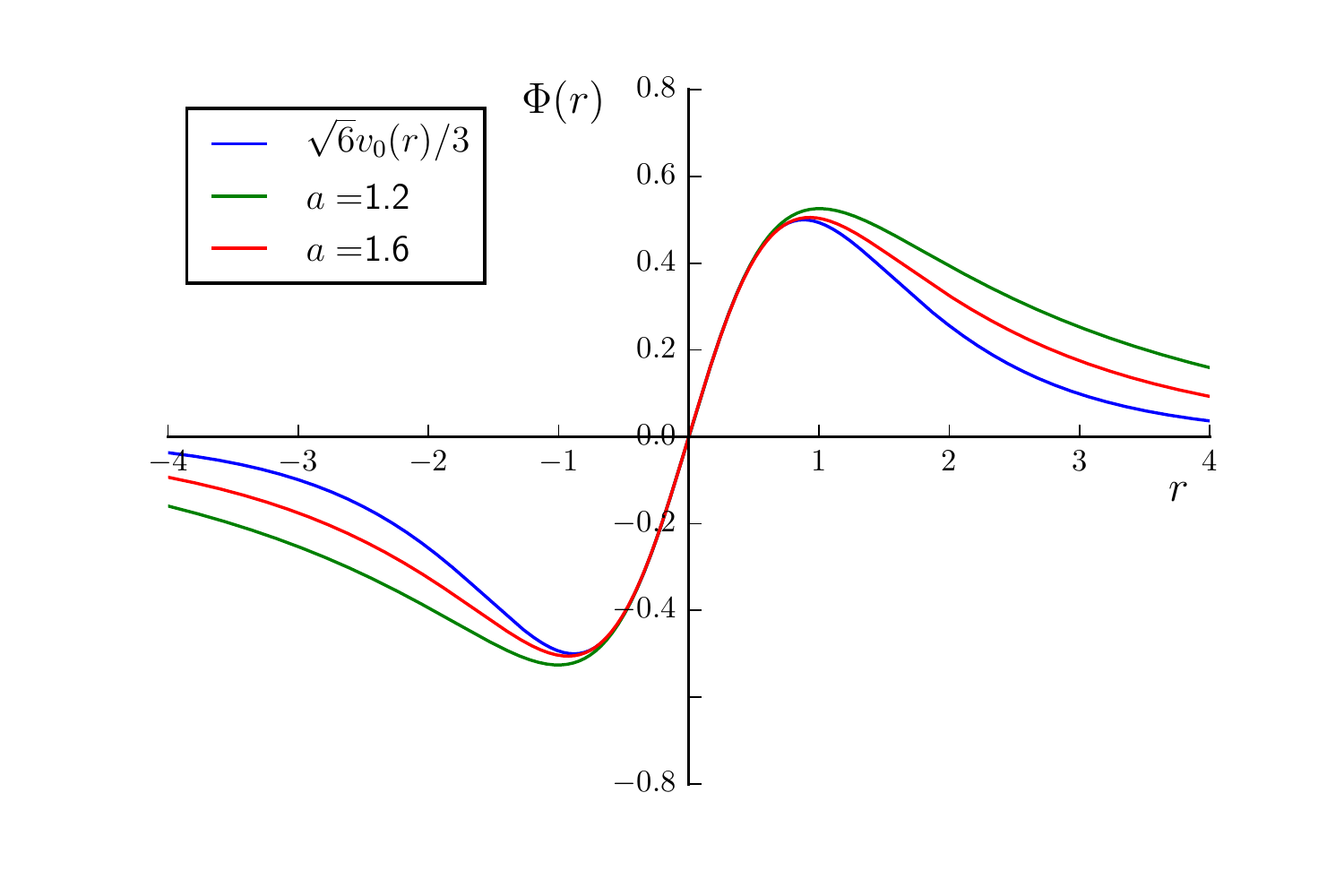}
\caption{\label{fig:odd_states} Odd bound states of the potential $V_0(r)$ and of the potential $V(r)$ for two different values of $a$.}
\end{figure}

\begin{figure}
\includegraphics[width=14cm]{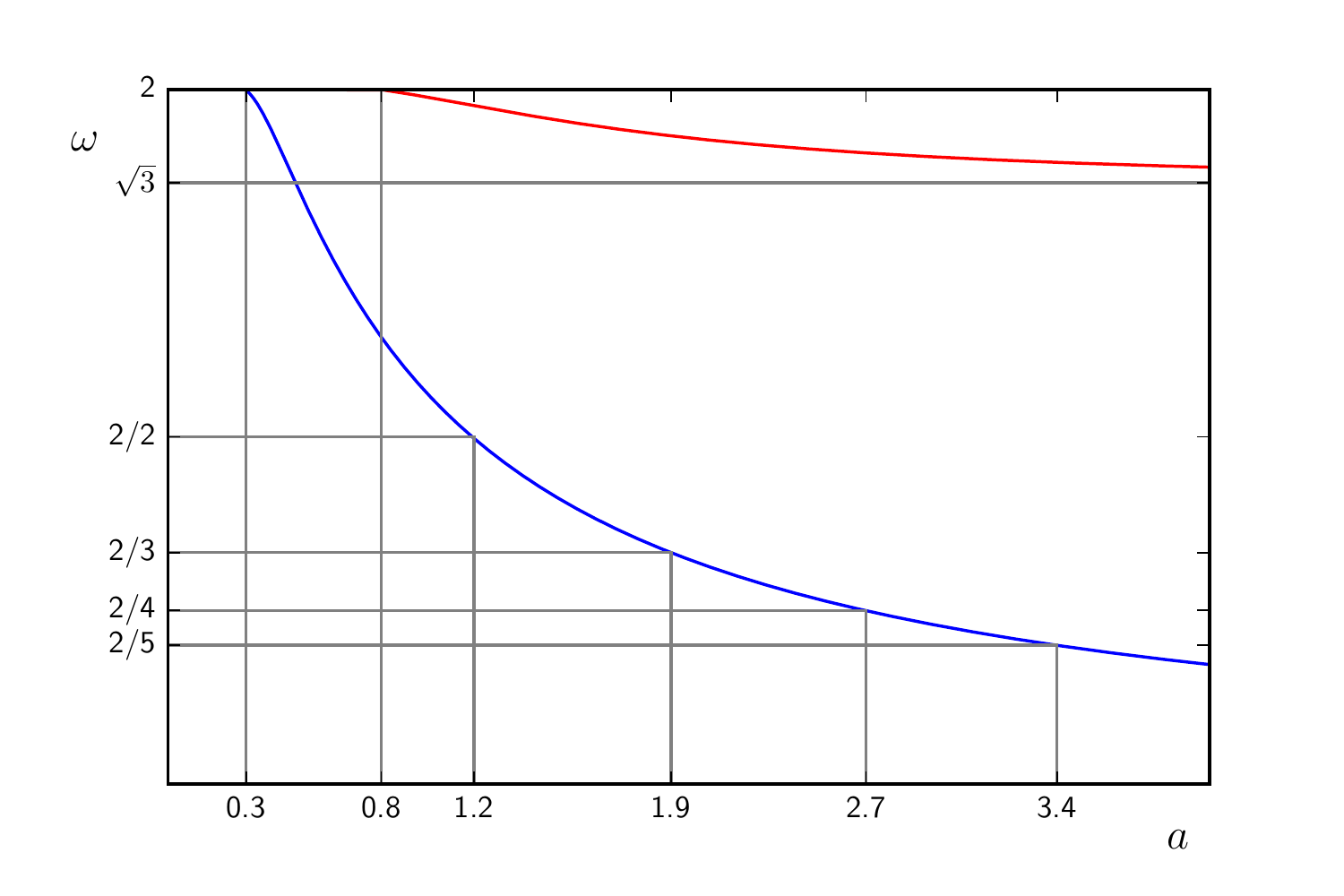}
\caption{\label{fig:omega_vs_a} The frequencies of the internal modes of the kink plotted against the wormhole radius $a$. The choice of axis ticks will be motivated in section \ref{sec:dynamics}.}
\end{figure}

For large $a$, the potential has both an even and an odd bound state which look qualitatively similar to the internal modes (\ref{eq:flat_vib_modes}) of the $\phi^4$ kink on $\mathbb{R}^{1,1}$. The bound states for several values of $a$ can be found in figures \ref{fig:even_states} and \ref{fig:odd_states}. As $a$ decreases, the eigenvalues $\omega^2$ of the bound states increase, until they disappear into the continuous spectrum ($\omega^2>4$). This disappearance will be further discussed in section \ref{sec:a_crit}. Their frequencies are plotted against $a$ in figure \ref{fig:omega_vs_a}.

\subsection{Large $a$ limit}

We can also perturbatively expand the eigenvalues of the eigenvalue problem (\ref{schrod}). Consider solutions to (\ref{eom.coords}) of the form\footnote{Note that in section \ref{sec:linearised_pert} we considered perturbations $v(r)$ which differ from $v_\epsilon(r)$ by a factor of $(r^2+a^2)^{-1/2}$, since such perturbations are described by a Schr\"odinger problem. Here it will be simpler to remove this factor; however there is a one--to--one correspondence between $v(r)$ and $v_\epsilon(r)$.} $\phi_\epsilon(r)=\Phi_\epsilon(r)+\mathrm{e}^{i\omega t}v_\epsilon(r)$, where $v_\epsilon$ is small.
These satisfy
\be
\label{eq:perturbed_large_a}
v_\epsilon^{\prime\prime} + \frac{2r\epsilon^2}{\epsilon^2r^2+1}v_\epsilon^\prime + 2(1-3\Phi_\epsilon^2)v_\epsilon = -\omega^2_\epsilon v_\epsilon.
\ee

Let $(v_\epsilon,\omega^2_\epsilon)$ be a solution to (\ref{eq:perturbed_large_a}) with
\[
\omega^2_\epsilon = \omega^2_0 + \epsilon^2\xi + \mathcal{O}(\epsilon^4) \quad \mathrm{and} \quad
v_\epsilon(r) = v_0(r) + \epsilon^2v_1(r) + \mathcal{O}(\epsilon^4).
\]
Our aim will be to find $\xi$. Substituting into (\ref{eq:perturbed_large_a}), at zero order we obtain the equation (\ref{eq:zero_order}) which controls the linear stability analysis of the $\phi^4$ kink on $\mathbb{R}^{1,1}$.

\begin{figure}
\includegraphics[width=14cm]{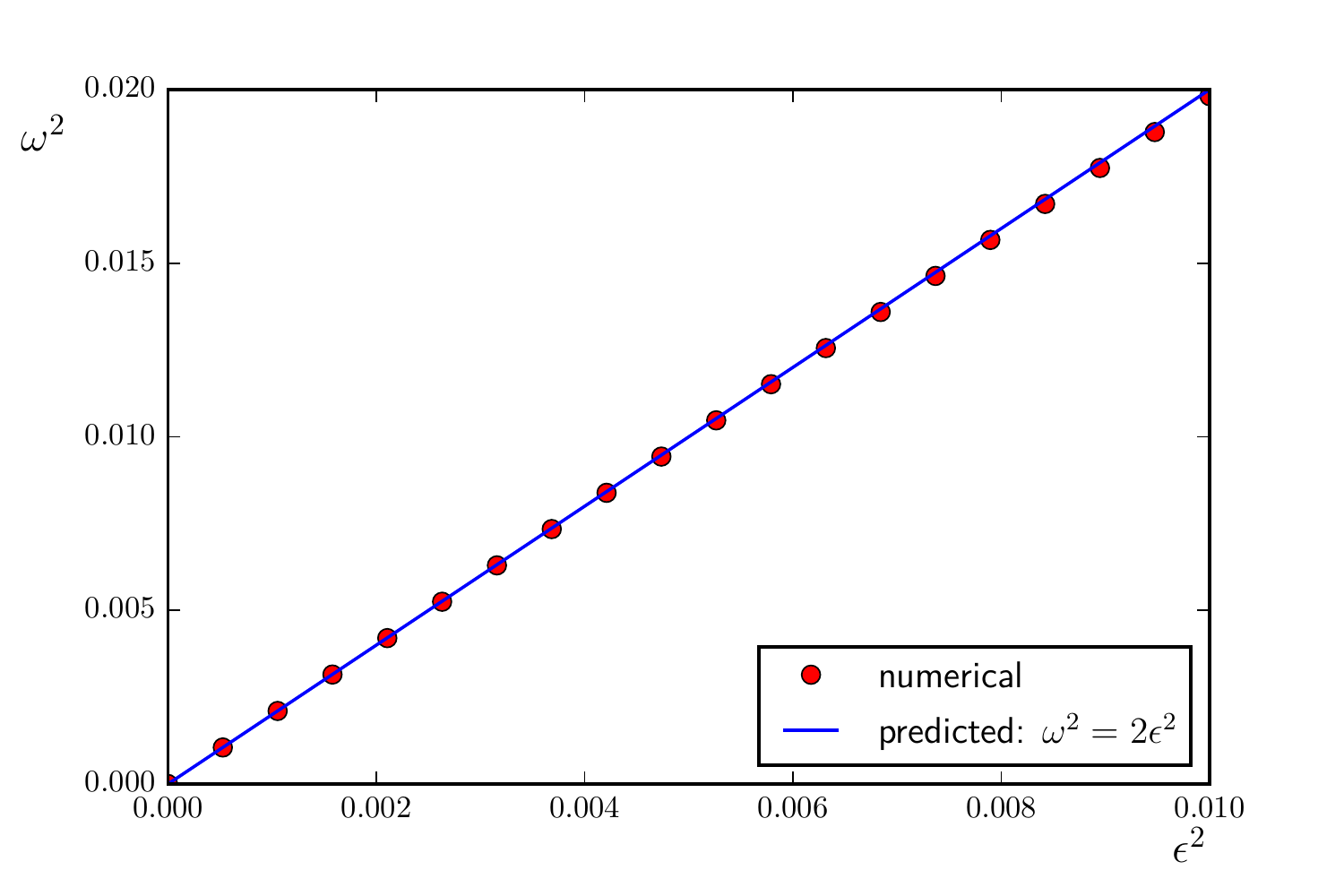}
\caption{\label{fig:xi_zero_mode}A comparison of the predicted and numerical calculations for the energy of the zero mode as a function of $\epsilon^2$ for small $\epsilon$. The numerical calculations were executed by finding the even bound states and their energies as described in section \ref{sec:finding_bound_states_numerically}.}
\end{figure}
\begin{figure}
\includegraphics[width=14cm]{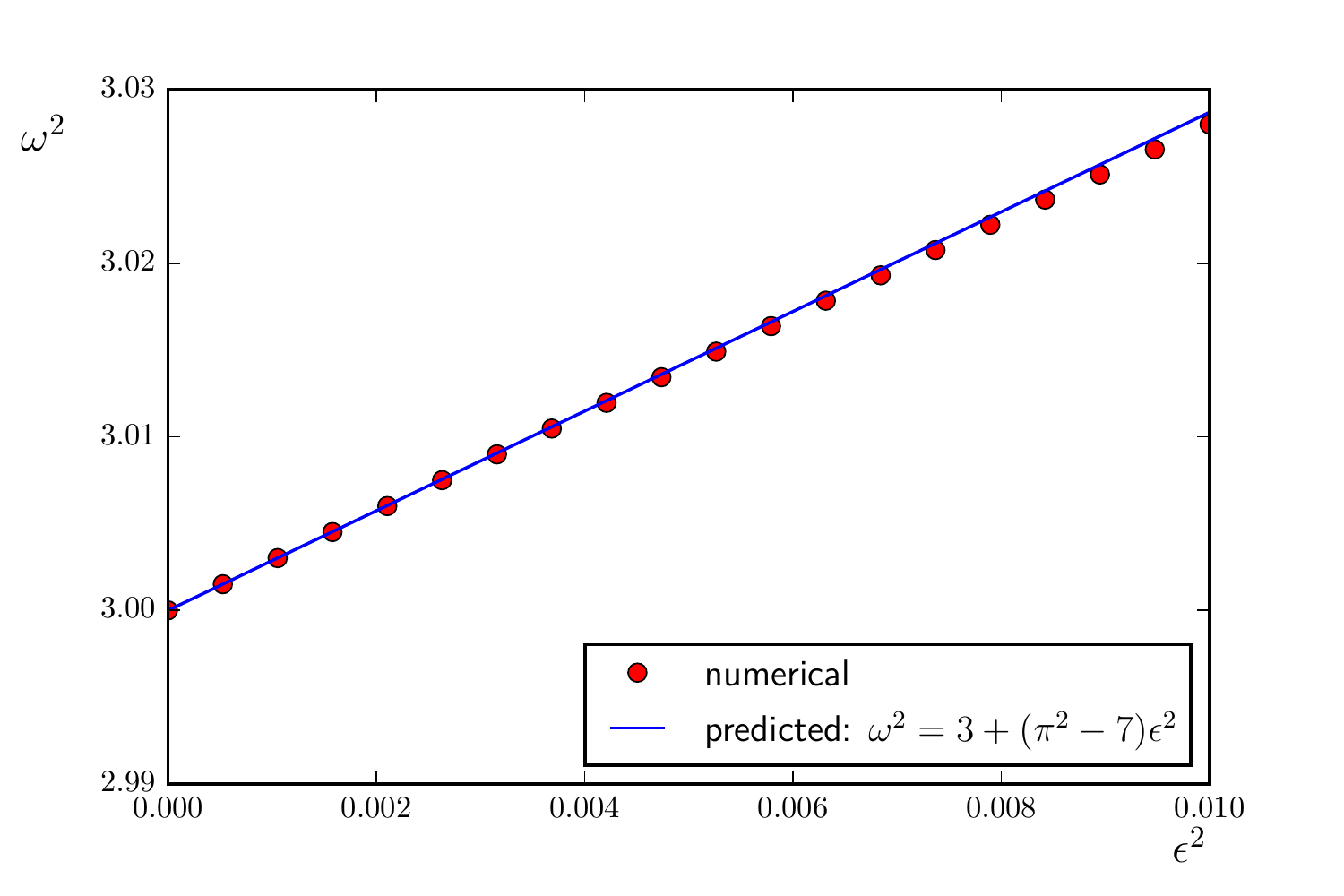}
\caption{\label{fig:xi_other_mode}A comparison of the predicted and numerical calculations for the energy of the odd vibrational mode as a function of $\epsilon^2$ for small $\epsilon$. The numerical calculations were executed by finding the odd bound states and their energies as described in section \ref{sec:finding_bound_states_numerically}.}
\end{figure}

The terms of order $\epsilon^2$ in (\ref{eq:perturbed_large_a}) give us
\be
\label{eq:first_order}
v_1^{\prime\prime} + 2rv_0^\prime + 2(1-3\Phi_0^2)v_1 - 12\Phi_0\Phi_1v_0 = -\omega^2_0v_1 - \xi v_0.
\ee
We multiply equation (\ref{eq:first_order}) by $v_0$, and subtract from this $v_1$ multiplied by equation (\ref{eq:zero_order}). Integrating the result from $r=-\infty$ to $r=\infty$, we find
\[
\int_{-\infty}^{\infty}\big(v_1^{\prime\prime}v_0-v_0^{\prime\prime}v_1\big)dr
+ \int_{-\infty}^{\infty}2rv_0^\prime v_0dr
- 12\int_{-\infty}^{\infty}\Phi_0\Phi_1v_0^2dr
=-\xi.
\]
In the first term the integrand is a total derivative, and the second term is easily found to be $-1$ using integration by parts. We thus obtain
\be
\label{eq:xi}
\xi = 1 + 12\int_{-\infty}^{\infty}\Phi_0\Phi_1v_0^2dr,
\ee
which we can evaluate for each of the solutions (\ref{eq:flat_vib_modes}) using the symbolic computation facility in Mathematica. We find $\xi=2$ in the case of the zero mode and $\xi=\pi^2-7$ in the case of the first non--trivial vibrational mode. We can check these values by finding $(v,\omega)$ numerically for a range of small values of $\epsilon$ and comparing $\omega^2$ to the $\omega^2_0+\xi\epsilon^2$ predicted here. The corresponding plots are shown in figures \ref{fig:xi_zero_mode} and \ref{fig:xi_other_mode}.

\subsection{Critical values of $a$} \label{sec:a_crit}

It is interesting to investigate the values of $a$ at which the internal modes disappear into the continuous spectrum. The larger of these, at which the odd internal mode disappears, we shall call $a_1$. The smaller one, at which the even internal mode disappears, we shall call $a_0$. 

The most convenient method of estimating $a_0$ and $a_1$ is based on the Sturm Oscillation Theorem \ref{th:sturm}. The points at which the even and odd internal modes disappear into the continuous spectrum are the points at which the zeros of the even and odd eigenfunctions of $L$ with $\omega^2=4$ disappear. We can thus examine the number of zeros of the odd eigenfunction with $\omega^2=4$ to determine the number of odd bound states with $\omega^2<4$. The critical value $a_1$ which we are searching for can then be found using a bisection method. An equivalent method using even bound states will yield an estimate of $a_0$.

One problem with this method is that we need the number of zeros in the interval $(0,\infty)$, and the shooting method we use to generate $\Phi(r)$ and $V(r)$ is only accurate up to a finite value of $r$. Since zeros of the eigenfunction with $\omega^2=4$ disappear at $r=\infty$, this limits the accuracy with which we can determine $a_0$ and $a_1$.

For the finite integration range which is accessible based on the shooting method, the odd state disappears at $a_1 \approx 0.8$ and the even state disappears at $a_0 \approx 0.3$.

It is well known that, for a potential $\mathcal{V}(r)$ which decays sufficiently quickly at the boundaries, the condition
\be
\label{eq:integral_condition}
I:=\int_{-\infty}^{\infty}\mathcal{V}(r)dr<0
\ee
is sufficient to ensure that the operator $-\p_r\p_r+\mathcal{V}(r)$ has at least one bound state. In fact, the condition $I\leq 0$ is sufficient \cite{BarrySimon}. However,  (\ref{eq:integral_condition}) is not a necessary condition: there are potentials which have at least one bound state where (\ref{eq:integral_condition}) is not satisfied. It is interesting to investigate the disappearance of our ground state in this context.

Note that $\mathcal{V}(r)$ must go to zero as $r\rightarrow\pm\infty$ to ensure that the integral converges, meaning that the relevant choice for us is $\mathcal{V}(r)=V(r)-4$. We then examine the value of this integral for the critical value $a=a_0$ when the ground state disappears. We find that $I\approx0$ at the critical value of $a_0 \approx 0.3$ given above. We can also search numerically for the value of $a$ at which $I=0$; this also occurs at around $a_0 \approx 0.3$. Thus our results would be consistent with the conjecture that (\ref{schrod}) has no bound states for $I>0$.

\section{Resonant Coupling of the Internal Modes to the Continuous Spectrum} \label{sec:dynamics}
We now move on to consider time dependent perturbations of the form
\[
\phi(t,r) = \Phi(r) + (r^2+a^2)^{-1/2}w(t,r),
\]
where we consider non--linear terms in $w(t,r)$. Substituting into (\ref{eom.coords}) we find
\be
\label{eq:nonsmall_pert}
w_{tt} = - Lw + f(w),
\ee
where we have defined
\be \label{eq:f}
f(w) = - \frac{6w^2\Phi}{\sqrt{a^2+r^2}} - \frac{2w^3}{a^2+r^2},
\ee
suppressing the dependence of $f$ on $r$ to simplify the notation. We will not have much need for the expression for $f$ other than to note that it contains terms which are quadratic and cubic in $w$.

If $a$ is large enough to allow internal modes, then these can only decay through resonant coupling to the continuous spectrum of $L$. The analogous process of decay to the $\phi^4$ kink on $\mathbb{R}^{1,1}$ was discussed in \cite{Manton&Merabet}, and the general theory was developed in \cite{SW98}. In the following sections we investigate this decay in the case of a single internal mode, before comparing our result with numerical data.

\subsection{Conjectured decay rate in the presence of a single internal mode} \label{sec:conjecture}
In this section we follow the analysis in \cite{SG}. Looking at figure \ref{fig:omega_vs_a}, we note that for $a\in(0.3,0.8)$ we have
\be \label{eq:spec_L}
\mathrm{spec}\ L=\{\omega^2\}\cap [m^2,\infty), \qquad \omega^2<m^2<4\omega^2
\ee
where $m^2=4$. As above, we denote the unique normalised eigenfunction of $L$ by $v$, so that $Lv=\omega^2v$. We will use $\langle\cdot,\cdot\rangle$ to denote the usual inner product on $\R$.

We decompose the perturbation as
\be
w(t,r) = \alpha(t)v(r) + \eta(t,r),
\ee
where $v(r)$ refers to the single even internal mode of the kink and $\eta$ is a superposition of states from the continuous spectrum of $L$. Where there is only one internal mode present, its frequency $\omega$ always lies in the upper half of the mass gap: $1<\omega<2$. This is important because it means that $2\omega$ lies within the continuous spectrum.

We substitute this into (\ref{eq:nonsmall_pert}) and project onto and away from the internal mode direction, obtaining the following equations for $\alpha$ and $\eta$:
\begin{align}
\ddot{\alpha} + \omega^2\alpha &= \langle v, f(\alpha v + \eta) \rangle \label{eq:alpha}\\
\ddot{\eta} + L\eta &= P^\perp f(\alpha v + \eta),\label{eq:eta}
\end{align}
where $P^\perp$ is the projection onto the space of eigenstates of $L$ which are orthogonal to $v$, given by
\be \label{eq:Pperp}
P^\perp\psi = \psi - \langle v,\psi\rangle v.
\ee

These equations have initial conditions $\alpha(0)$ and $\eta(0,r)$ such that
\begin{align*}
\phi(0,r) &= \Phi(r) + (r^2+a^2)^{-1/2}(\alpha(0)v(r) + \eta(0,r)), \quad \mathrm{and} \\
\dot{\phi}(0,r) &= (r^2+a^2)^{-1/2}(\dot{\alpha}(0)v(r) + \dot{\eta}(0,r)).
\end{align*}

In the following analysis we investigate decay of $\alpha(t)$. Equation (\ref{eq:alpha}) has a homogeneous solution consisting of oscillations with frequency $\omega$. Since $2\omega$ lies within the continuous spectrum of $L$, there will be a resonant interaction between the these oscillations and the radiation modes in $\eta$ with frequencies $\pm 2 \omega$, arising from the term of order $\alpha^2$ in the RHS of (\ref{eq:eta}). Thus, to leading order, (\ref{eq:eta}) is a driven wave equation with driving frequency $2\omega$. This resonant part of $\eta$ will have a back--reaction on $\alpha$ through (\ref{eq:alpha}), which will result in decay of the internal mode oscillations.

We now define $\alpha_1=\alpha,\ \omega\alpha_2=\dot{\alpha}_1$  so that (\ref{eq:alpha}) becomes
\be
\begin{cases*} \nonumber
\dot{\alpha}_1 = \omega\alpha_2, \\
\dot{\alpha}_2 = - \omega\alpha_1 + \frac{1}{\omega}\langle v, f(\alpha_1 v+\eta)\rangle,
\end{cases*}
\ee
or equivalently
\be \label{eq:A}
\dot{A} = -i\omega A + \frac{i}{\omega}\bigg< v,f\bigg(\frac{1}{2}(A + \bar{A})v+\eta\bigg)\bigg>,
\ee
where $A=\alpha_1 + i\alpha_2$. Next we write $\eta_1=\eta,\ \eta_2=\dot{\eta}_1$, converting (\ref{eq:eta}) to
\be \label{eq:eta2}
\begin{cases*}
\dot{\eta}_1 =\eta_2, \\
\dot{\eta}_2 = -L\eta_1 + P^\perp f\big( \frac{1}{2} (A+\bar{A}) v + \eta_1\big).
\end{cases*}
\ee

We will regard the right hand sides of (\ref{eq:A}) and (\ref{eq:eta2}) as power series in $A$ and $\eta$. Terms which we expect to be higher order will not be treated rigorously; for this reason, our analysis will produce only a conjecture about the decay rate. Numerical evidence concerning the conjecture will be discussed in section \ref{sec:numerics}.

It will be helpful to introduce the notation $\cO_p(A,\eta)$ to mean terms of at least order $p$ in $A,\bar{A},\eta_1,\eta_2$, so that $A^2$, $\eta_1^2 $ and $\bar{A}\eta_1$ are all examples of terms which are $\cO_2(A,\eta)$. Currently, the coupling between (\ref{eq:A}) and (\ref{eq:eta2}) is $\cO_2(A,\eta)$. We will write
\[
f\bigg( \frac{1}{2} (A+\bar{A}) v + \eta_1\bigg) = \sum_{k+l\geq 2} f_{kl}A^k\bar{A}^l + \sum_{\substack{k+l\geq 1 \\ n\geq 1}} f_{kln}\eta_1A^k\bar{A}^l
\]
where $k,l,n$ are non--negative, to elucidate the lowest order terms in (\ref{eq:eta2}). Note that $f_{kl}$ and $f_{kln}$ are decaying functions of $r$ defined by (\ref{eq:f}). We can then write
\[
P^\perp\bigg[ f\bigg( \frac{1}{2} (A+\bar{A}) v + \eta_1\bigg) \bigg] = \sum_{k+l=2} P^\perp[f_{kl}]A^k\bar{A}^l + \sum_{k+l=1} P^\perp [f_{kl1}\eta_1]A^k\bar{A}^l + \cO_3(A,\eta).
\]

Terms in (\ref{eq:A}) with imaginary coefficients correspond to rotation in the complex plane, and thus to oscillatory behaviour in $\alpha$. At first order, $A$ oscillates with frequency $\omega$. This is exactly the behaviour expected in the linearised theory discussed in section \ref{sec:linearised_pert}. In fact, a priori, all the terms in the power series for $\dot{A}$ have coefficients which are purely imaginary.

The next step in our analysis will be to attempt a change of variable $\eta_i\mapsto\tilde{\eta}_i$ in (\ref{eq:eta2}) so that its right hand side is $\cO_3(A,\tilde{\eta})$, meaning $\tilde{\eta}$ is $\cO(A^3)$. It will turn out that the required change of variables is complex. The result will be a term in (\ref{eq:A}) which is $\cO(A^3)$ and has a real coefficient. This will be the lowest order term with a real coefficient, and thus the key resonant damping term. 

We write the change of variables as
\begin{equation} \label{eq:eta_change}
\eta_1 = \tilde{\eta}_1 + \sum_{k+l=2}b_{kl}A^k\bar{A}^l, \qquad \eta_2 = \tilde{\eta}_2 + \sum_{k+l=2}c_{kl}A^k\bar{A}^l,
\end{equation}
where $b_{kl}$ and $c_{kl}$ are functions of $r$ which are so far undetermined. Differentiating with respect to time and using (\ref{eq:A}), we find
\begin{align*}
\dot{\eta}_1 &= \dot{\tilde{\eta}}_1 - i\omega\sum_{k+l=2}b_{kl}(k-l)A^k\bar{A}^l+\cO_3(A,\tilde{\eta}), \\[5pt]
 \dot{\eta}_2 &= \dot{\tilde{\eta}}_2 - i\omega\sum_{k+l=2}c_{kl}(k-l)A^k\bar{A}^l + \cO_3(A,\tilde{\eta}).
\end{align*}
We equate these to the right hand sides of (\ref{eq:eta2}), substituting from (\ref{eq:eta_change}) and requiring that
\be \label{eq:eta_tilde}
\dot{\tilde{\eta}}_1 = \tilde{\eta}_2 + \cO_3(A,\tilde{\eta}), \qquad \dot{\tilde{\eta}}_2 = -L\tilde{\eta}_1 + \cO_3(A,\tilde{\eta}).
\ee
This yields
\[
-i\omega b_{kl}(k-l)=c_{kl} \qquad\mbox{and}\qquad -i\omega c_{kl}(k-l)=-Lb_{kl} + P^\perp[f_{kl}]
\]
for $k+l=2$, where we have discarded
\begin{align*}
\sum_{k+l=1} P^\perp[f_{kl1}\eta_1]A^k\bar{A}^l &= \sum_{k+l=1} P^\perp [f_{kl1}\tilde{\eta}_1]A^k\bar{A}^l + \sum_{\substack{k+l=1 \\ p+q=2}} P^\perp [f_{kl1}b_{pq}]A^{k+p}\bar{A}^{l+q} \\
&= \cO_3(A)
\end{align*}
because $\tilde{\eta}$ is at least third order in $A$.

The change of variables (\ref{eq:eta_change}) is now given by the solution to
\be \label{eq:solve_for_b}
\big(L-\omega^2(k-l)^2\big)b_{kl}=P^\perp[f_{kl}].
\ee
Because of the spectrum of $L$ given in (\ref{eq:spec_L}), for $(k,l)\in(2,0)\cup (0,2)$ the solution $b_{kl}$ is in general a complex function of $r$, whilst for $k=l=1$ the solution is real and decaying. The reason for this can be understood using the variation of parameters method for inhomogeneous ODEs.

Let $g(r)$ be such that $\langle g,g \rangle$ is finite, and $\lambda\geq 0$ a constant. The general solution of
\[
(L-\lambda^2)b(r)=g(r)
\]
is given by
\[
b(r)=Z_2(r)\int_{-\infty}^r \frac{1}{W(r^\prime)}Z_1(r^\prime)g(r^\prime)dr^\prime + Z_1(r)\int_r^\infty \frac{1}{W(r^\prime)} Z_2(r^\prime)g(r^\prime) dr^\prime,
\]
where $\{Z_1,Z_2\}$ is a basis for solutions to the homogeneous equation with Wronskian $W(r)=Z_1Z_2^\prime -Z_2Z_1^\prime$. The basis must be chosen so that the above integrals converge. 

For $k=l=1$, so that $\lambda^2=0$ and hence $\lambda^2<m^2$, we can choose a basis such that $W=1$ and $Z_1,Z_2$ are both real, and they decay to zero in the limits $r\rightarrow -\infty$ and $r\rightarrow\infty$ respectively. Then
\be \label{eq:decaying_soln}
b_{11}(r) = Z_2(r)\int_{-\infty}^r Z_1(r^\prime)P^\perp[f_{11}](r^\prime)dr^\prime + Z_1(r)\int_r^{\infty} Z_2(r^\prime)P^\perp [f_{11}](r^\prime) dr^\prime.
\ee
For $\lambda^2\geq m^2$, we cannot choose a real solution in general. In the case $(k,l)\in(2,0)\cup (0,2)$, we take as a basis the Jost functions $\{j_\pm\}$, defined by
\begin{equation*}
j_\pm(r) \sim e^{\pm i\xi r} \ \mbox{as} \ r\rightarrow\infty,
\end{equation*}
where $\xi=\sqrt{4\omega^2-m^2}$. Their Wronskian is then $W(j_+,j_-)=-2i\xi$, and we write the solution as
\be \label{eq:oscillating_soln}
b_{02}(r) = b_{20}(r) = \frac{ij_-(r)}{2\xi}\int_{-\infty}^r j_+(r^\prime)P^\perp[f_{20}](r^\prime)dr^\prime + \frac{ij_+(r)}{2\xi}\int_r^{\infty} j_-(r^\prime)P^\perp[f_{20}](r^\prime) dr^\prime.
\ee

Finally, we use (\ref{eq:decaying_soln}) and (\ref{eq:oscillating_soln}) to change variable $\eta_i\mapsto\tilde{\eta}_i$ in (\ref{eq:A}), obtaining
\be \label{eq:A2}
\dot{A} = -i\omega A + \frac{i}{\omega}\Bigg(\sum_{2\leq k+l\leq 3}\langle v, f_{kl}\rangle A^k\bar{A}^l + \sum_{\substack{k+l=1 \\ p+q=2}} \langle v, f_{kl1}b_{pq}\rangle A^{k+p}\bar{A}^{l+q} + \cO(A^4)\Bigg),
\ee
where we have ignored terms containing $\tilde{\eta}_1$ since these are at least fourth order in $A$. We can now see that, of the terms which we have written explicitly, the only ones that can give a real contribution to $\dot{A}$ are those containing $b_{02}$ and $b_{20}$. We thus find
\begin{align*}
\frac{d}{dt}|A|^2 &= \dot{A}\bar{A} + A\dot{\bar{A}} = 2\Real[\dot{A}\bar{A}] \\
&= \frac{2}{\omega} \Real\Bigg[i\sum_{k+l=1}\bigg( \langle v, f_{kl1}b_{20}\rangle A^{k+2}\bar{A}^{l+1} + \langle v, f_{kl1}b_{02}\rangle A^{k}\bar{A}^{l+3} \bigg) \Bigg] + \cO(A^5) \\
&= -\frac{2}{\omega} \mathbb{I}\mathrm{m}\bigg[\langle v, f_{101}b_{20}\rangle(A^3\bar{A} + A^2\bar{A}^2 + A\bar{A}^3 + \bar{A}^4) \bigg] + \cO(A^5).
\end{align*}
In particular, the term $A^2\bar{A}^2=|A|^4$ is real and non--oscillating, giving a contribution
\[
\frac{d}{dt}|A|^2 \sim -\frac{2}{\omega} \mathbb{I}\mathrm{m}\big[\langle v, f_{101}b_{20}\rangle\big]|A|^4.
\]
The terms $A^3\bar{A}$, $A\bar{A}^3$ and $\bar{A}^4$, on the other hand, would be expected to oscillate at frequencies $2\omega$ and $4\omega$ at first order, and thus time average to zero.

Hence we conclude
\be \label{eq:decay_behaviou}
|A|\sim \bigg(\Gamma t + \frac{1}{|A(0)|^2}\bigg)^{-1/2}, \qquad \Gamma:=\frac{2}{\omega} \mathbb{I}\mathrm{m}\big[\langle v, f_{101}b_{20}\rangle\big].
\ee

The constant $\Gamma$ is a function of $a$ which can be calculated explicitly. Using (\ref{eq:f}) and (\ref{eq:oscillating_soln}) gives
\begin{align*}
\langle v, f_{101}b_{20}\rangle = \int_{-\infty}^{\infty}dr\ v(r)f_{101}(r)\bigg(&\frac{ij_-(r)}{2\xi}\int_{-\infty}^r j_+(r^\prime)P^\perp[f_{20}](r^\prime)dr^\prime \\
&+ \frac{ij_+(r)}{2\xi}\int_r^{\infty} j_-(r^\prime)P^\perp[f_{20}](r^\prime) dr^\prime\bigg).
\end{align*}
We now use the facts that $f_{20}=vf_{101}/4$, and $P^\perp[f_{20}]=f_{20}-\langle v,f_{20}\rangle v$. Note that $f_{20}$ is an odd function of $r$, so in fact $\langle v,f_{20}\rangle=0$ and so $P^\perp[f_{20}]=f_{20}$. We thus obtain
\begin{align*}
\langle v, f_{101}b_{20}\rangle = \frac{2i}{\xi}\bigg(&\int_{-\infty}^{\infty}dr
\int_{-\infty}^r dr^\prime f_{20}(r)j_-(r)f_{20}(r^\prime)j_+(r^\prime) \\
&+ \int_{-\infty}^{\infty}dr\int_r^{\infty}dr^\prime f_{20}(r)j_+(r)f_{20}(r^\prime)j_-(r^\prime)\bigg).
\end{align*}
The two double integrals are integrals over complementary halves of the $(r,r^\prime)$ plane, and thus sum to a single integral over the full plane. Hence
\[
\langle v, f_{101}b_{20}\rangle = \frac{2i}{\xi}\int_{-\infty}^{\infty}f_{20}(r)j_+(r)dr
\int_{-\infty}^\infty  f_{20}(r^\prime)j_-(r^\prime) dr^\prime=\frac{2i}{\xi}|\langle f_{20}, j_+ \rangle|^2,
\]
since $j_\pm$ are complex conjugates.

Combining this with (\ref{eq:decay_behaviou}) gives
\[
\Gamma = \frac{4}{\omega\xi}|\langle f_{20}, j_+\rangle|^2.
\]
The so--called Fermi Golden Rule then reads
\[
|\langle f_{20}, j_+\rangle|\neq 0.
\]

\subsection{Numerical investigation of the conjectured decay rate}
\label{sec:numerics}

\begin{figure}
\includegraphics[width=14cm]{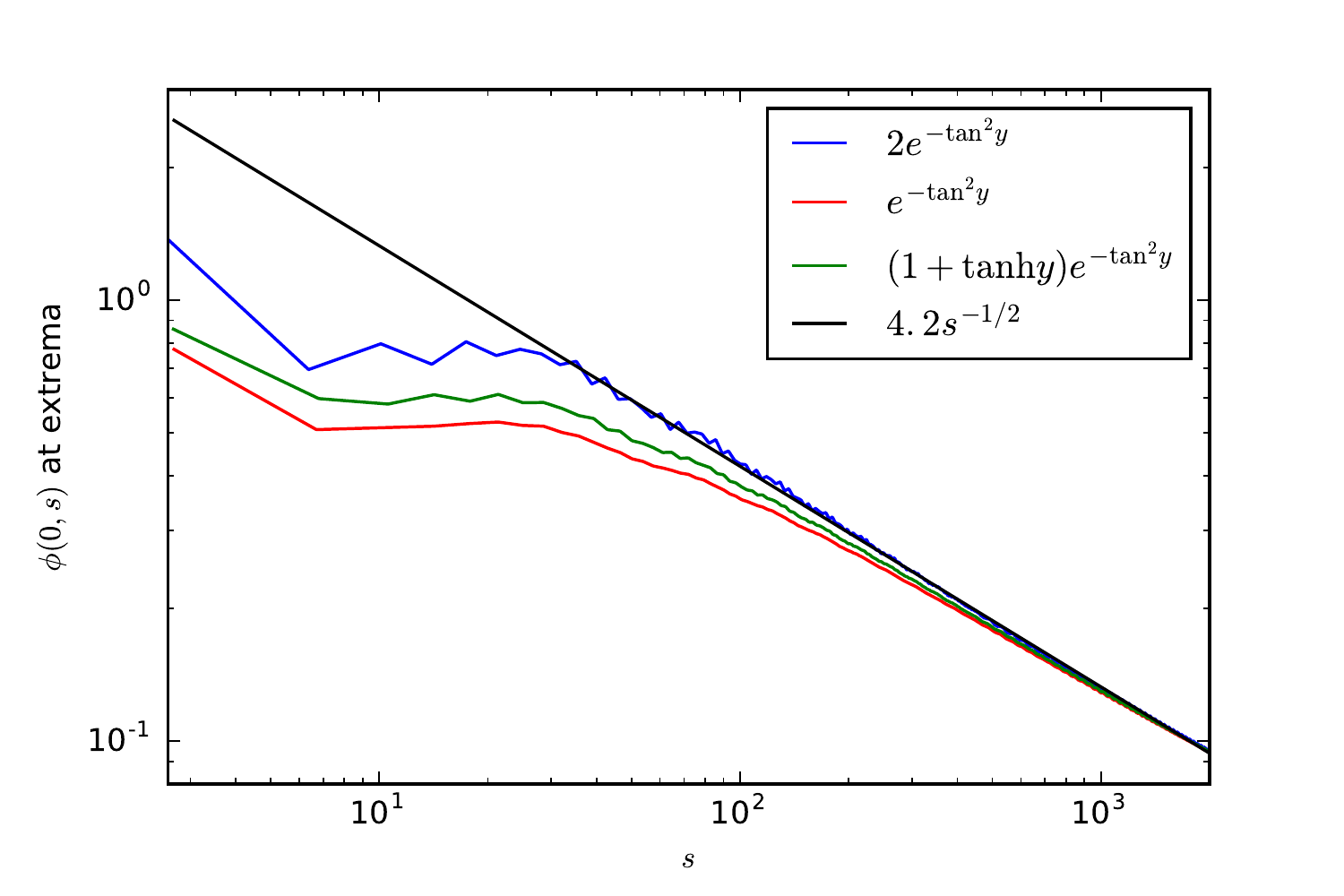}
\caption{\label{fig:decay} The decay of internal mode oscillations for various initial conditions when $a=0.5$. Note that $\phi(0,s)$ is used as a proxy for the internal mode amplitude, and we use a log--log scale to elucidate the dependence on $s^{-1/2}$ in the large $s$ limit. The lines are labelled in the legend by the initial conditions which produced them, with the exception of the gradient line $4.2s^{-1/2}$.}
\end{figure}

In order to integrate the PDE (\ref{eom.coords}) to large times $t$, we employ the method of hyperboloidal foliations and scri--fixing \cite{hyp_foliations}. Following \cite{wavemaps, SG}, we define
\[
s = \frac{t}{a} - \sqrt{\frac{r^2}{a^2}+1}, \qquad y=\mathrm{arctan}\Big(\frac{r}{a}\Big),
\]
resulting in a hyperbolic equation
\be
\label{eq:s_and_y}
\p_s\p_sF + 2\mathrm{sin}(y)\p_y\p_sF + \frac{1 + \mathrm{sin}^2(y)}{\mathrm{cos}(y)}\p_sF = \mathrm{cos}^2(y)\p_y\p_yF + 2a^2\frac{F(1-F^2)}{\mathrm{cos}^2(y)}.
\ee
for $F(s,y)=\phi(t,r)$. 

We solve the corresponding initial value problem at space--like hypersurfaces of constant $s$, specifying $\phi(s=0,y)$ and $\p_s\phi(s=0,y)$. No boundary conditions are required, since the principal symbol of (\ref{eq:s_and_y}) degenerates to $\p_s(\p_s\pm 2\p_y)$ as $y\rightarrow\pm \pi/2$, so there are no ingoing characteristics. This reflects the fact that no information comes in from future null infinity.

Following \cite{wavemaps, sym_hyp} we define the auxiliary variables
\[
\Psi = \p_yF, \qquad \Pi = \p_sF + \mathrm{sin}y\p_yF
\]
to obtain the first order symmetric hyperbolic system
\begin{align}
\p_sF &= \Pi - \Psi\mathrm{sin}y \\
\label{eq:Psi} \p_s\Psi &= \p_y(\Pi-\Psi\mathrm{sin}y) \\
\p_s\Pi &= \p_y(\Psi-\Pi\mathrm{sin}y) + 2\mathrm{tan}y(\Psi-\Pi\mathrm{sin}y) + 2a^2\frac{F(1-F^2)}{\mathrm{cos}^2y},
\end{align}
which we solve numerically using the method of lines. Kreiss--Oliger dissipation is required to reduce unphysical high--frequency noise. We also add the term $-0.1(\Psi-\p_yF)$ to the right hand side of equation (\ref{eq:Psi}) to suppress violation of the constraint $\Psi=\p_yF$.

We are interested in the range of values $a_0<a<a_1$ for which the kink has exactly one internal mode. We find that, for fixed but arbitrary $y$, $F(s,y)$ oscillates in $s$ with a frequency close to the internal mode frequency, and that these oscillations tend towards a decay rate of $s^{-1/2}$, as we expect from section \ref{sec:conjecture}. Plots demonstrating this decay at $y=0$ for $a=0.5$ are shown in figure \ref{fig:decay}. Note that the constant $4.2$ is related to $\Gamma$ as defined in (\ref{eq:decay_behaviou}).

\subsection{Expected decay rates in other regimes}
From figure \ref{fig:omega_vs_a}, we see that the second internal mode appears before the frequency of the first internal mode moves out of the range $(m/2,m)$. In the presence of more than one internal mode, we expect complicated coupling between their amplitudes, making the behaviour at large times very difficult to predict. However, if we restrict to odd initial data, the solution to (\ref{eq:nonsmall_pert}) remains odd. This means that the even internal mode can never be excited, so the system effectively has only one internal mode. In this case, the analysis in section \ref{sec:conjecture} still applies, since the frequency of the odd internal mode always lies in the range $(m/2,m)$, so we expect its amplitude to decay like $s^{-1/2}$.

For general initial data and wormhole radii $a>a_1$, we cannot produce a concrete conjecture about the decay rate. However, we expect the behaviour of the system to depend on the locations of the two internal mode frequencies within the mass gap. The analysis in section \ref{sec:conjecture} for a single internal mode suggests that for frequencies less than $m/2$, a real contribution to $\dot{A}$ does not appear until at least $\mathcal{O}(A^4)$. In this case, we require another change of variable in the radiation equation to rule out the contribution of the radiative term at higher order. We then expect to solve an equivalent of (\ref{eq:solve_for_b}) where $k+l=3$ to find the required change of variable. If $9\omega^2<m^2$ so that the solution is still real, we can proceed by induction, changing the variable until the solution is complex. A real contribution to (\ref{eq:A2}) will only be obtained for $k+l=N$ such that $N^2\omega^2>m^2$. This would mean a real contribution to $\dot{A}$ at $\mathcal{O}(A^{N+1})$, and hence result in a decay rate of $s^{-1/N}$. Further detail can be found in \cite{SG}.

Although the presence of a second internal mode complicates the dynamics, we still expect the smallest $N$ such that the even internal mode frequency $\omega$ satisfies $N^2\omega^2>m^2$ to be an important factor in the behaviour at large $s$. The axis ticks in figure \ref{fig:omega_vs_a} show the value of $N$ for a range of $a$.

\section{Summary and Discussion}

We have found that the modified kink is topologically and linearly stable, and investigated its asymptotic stability for the range of $a$ where exactly one discrete mode is present. It would be interesting to expand the investigation in section \ref{sec:dynamics} to the case when both discrete modes are present. This problem is much more complicated because of the extra terms in (\ref{eq:eta}) and (\ref{eq:alpha}) coming from the amplitude of the second internal mode. Similar problems have been discussed in \cite{Weinstein}, although no such analysis has been done for non--linear Klein--Gordon equation of this type with two discrete modes. The $\phi^4$ theory on the wormhole presents a useful setting to undertake such analysis because the kink has exactly two discrete modes for any $a>a_1$, and because their frequencies can be controlled by the parameter $a$.

This model shares an interesting property with its sine--Gordon counterpart in that we expect a discontinuous change in decay behaviour when $a$ moves out of the range $a_0<a<a_1$. Insight from the $\phi^4$ case may help to elucidate the character of such discontinuous changes.

{\bf Acknowledgements.} The author would like to thank Maciej Dunajski, Piotr Bizo\'n and Michal Kahl for helpful discussions.


\begin{thebibliography}{}

\bibitem{wavemaps} Bizo\'n, P. and Kahl, M. (2015)
Wave maps on a wormhole. Phys. Rev. D {\bf 91} 065003

\bibitem{SG} Bizo\'n, P., Dunajski, M., Kahl, M. and Kowalczyk, M. (2019) Sine--Gordon on a Wormhole (preprint).

\bibitem{Bronnikov} Bronnikov, K. A. (1973)
Scalar--tensor theory and scalar charge. Acta Phys. Polonica {\bf b4} 251--266

\bibitem{Derrick} Derrick, G. H. (1964)
Comments on nonlinear wave equations as models for elementary particles.
J. Math. Phys. {\bf 5}, 1252

\bibitem{Ellis} Ellis, H. G. (1973)
Ether flow through a drainhole: A particle model in general relativity. J. Math. Phys. {\bf 14}, 104--118

\bibitem{Hall} Hall, B. C. (2013)
Quantum Theory for Mathematicians. Graduate Texts in Mathematics {\bf 267}, Springer

\bibitem{KowalczykMM} Kowalczyk, M., Martel, Y. and Mu\~noz, C.
Kink Dynamics in the $\phi^4$ Model: Asymptotic Stability for Odd Perturbations in the Energy Space. 

\bibitem{Manton&Merabet} Manton, N. S. and Merabet, H. (1997)
$\phi^4$ Kinks - Gradient Flow and Dynamics. Nonlinearity {\bf 10} 3--18

\bibitem{Manton&Sutcliffe} Manton, N. S. and Sutcliffe, P. (2004)
Topological Solitons. Cambridge Monographs on Mathematical Physical, Cambridge University Press, Cambridge

\bibitem{Segur} Segur, H. (1983)
Wobbling kinks in $\varphi^4$ and sine-Gordon theory. Journal of Mathematical Physics {\bf 24}, 1439--1444

\bibitem{BarrySimon} Simon, B. (1975)
The Bound State of Weakly Coupled Schr\"odinger Operators in One and Two Dimensions. Annals of Physics {\bf 97} 279--288

\bibitem{SW98} Soffer, A. and Weinstein, M. I. (1998)
Time Dependent Resonance Theory. Geometric and Functional Analysis {\bf 8} 1086--1128

\bibitem{SW99} Soffer, A. and Weinstein, M. I. (1999)
Resonances, radiation damping and instability in Hamiltonian nonlinear wave equations. Inventiones Mathematicae {\bf 136} 9--74

\bibitem{TT} Tao, T. (2009)
Why are Solitons Stable? American Mathematical Society {\bf 46} 1-33

\bibitem{Weinstein} (2002)
Extended Hamiltonian Systems. Handbook of Dynamical Systems IB, 1135--1153

\bibitem{XinliangSoffer} Xinliang, An. and Soffer, A. (2017)
Fermi's golden rule and $H^1$ scattering for nonlinear Klein--Gordon equations with metastable states

\bibitem{hyp_foliations} Zenginoglu, A. (2008)
Hyperboloidal foliations and scri-fixing. Classical and Quantum Gravity {\bf 25} 145002

\bibitem{sym_hyp} Zenginoglu, A. (2008)
A hyperboloidal study of tail decay rates for scalar and Yang-Mills fields. Classical and Quantum Gravity {\bf 25} 175013






\end{thebibliography}
\end{document}